
\magnification \magstep1

\footline={\hfil}
\rightline{DAMTP preprint R93/25}
\rightline{gr-qc/9310002}
\vskip 0.7cm

\centerline{ \bf Anti-Gravity Bounds and the Ricci Tensor}
\vskip 0.7cm
\centerline { G W Gibbons\footnote{$^1$}{e-mail address GWG1@AMTP.CAM.AC.UK}
and C G Wells\footnote{$^2$}{e-mail address CGW11@AMTP.CAM.AC.UK}}
\vskip 0.7cm
\centerline { DAMTP}
\vskip 0.7cm
\centerline {University of Cambridge}
\vskip 0.7cm
\centerline {Silver Street}
\vskip 0.7cm
\centerline {Cambridge}
\vskip 0.7cm
\centerline {CB3 9EW}
\vskip 0.7cm

{
\narrower
{\bf Abstract}

\narrower
\noindent Recently Penrose, Sorkin and Woolgar have developed a new technique
for proving
the positive mass theorem in general relativity. We extend their result to
produce a new inequality relating the mass, electric and scalar charges in
theories coupling to a dilaton in the usual way. Using a five dimensional
formalism, our result provides new
information not available from the existing techniques. The main result may be
simply expressed as
$$
 M+g\Sigma\ge{\sqrt{1+g^2}|Q|\over\sqrt {4\pi G}}\ .
$$

\noindent where $M$ is the A.D.M. mass, $Q$ is the electric charge, $\Sigma$
the scalar
charge and $g$ is the dilaton coupling parameter.

}

\vskip 2 cm
\centerline{\sl Submitted to Comm. Math. Phys.}
\vfill
\eject
\count0=1
\footline={\hss\tenrm\folio\hss}
\headline{\centerline{ANTI-GRAVITY BOUNDS AND THE RICCI TENSOR}}

\beginsection 1. Introduction

 In theories of gravity in four
spacetime dimensions in which, in addition to the
graviton, there are additional massless boson fields of spin zero  and
spin one the long range inverse square law attraction produced by the
graviton
and scalar or pseudo-scalar particles can, to some extent,
be compensated by the long range repulsion produced by the spin
one vector fields.
However if one insists that there are no naked singularities
and that the sources, if there are any,
satisfy appropriate conditions, one typically finds, for
example by studying black hole solutions of the equations of motion, that
the repulsive forces can at best exactly compensate the attractive forces
to produce a state of equipoise, they can never overwhelm the
attractive forces altogether.

This phenomenon is sometimes referred to as
\lq \lq anti-gravity\rq\rq and it arises in various theories, including
supergravity and Kaluza-Klein theories. It is possible
to investigate anti-gravity using some ideas from supersymmetry and
supergravity even though the theories one may be interested
in are not necessarily supergravity theories. Using a generalisation [1]
of Witten's proof of the positive mass theorem [2] it is possible
to show that the absence of systems which are repulsive is a general
phenomenon because the total mass of an isolated system is bounded
below, in suitable units,  by the magnitude of any of its central charges.
A state of anti-gravity may occur if this bound is attained. In
supersymmetric theories this state is
typically supersymmetric.

The aim of this paper is to place what seem to be different
limits on the ADM mass $M$,
scalar charge $\Sigma$ and electric charge $Q$ of an isolated system
which also guarantees that it attracts at large distances from
a different view-point. The idea is to adapt the technique recently
exploited by
Penrose, Sorkin and Woolgar to prove a version of the
Positive Mass Theorem [3]. These authors establish a close link between
the attractive properties of an isolated source (i.e. positivity of the
total ADM mass $M$) and the non-negativity of the
Ricci tensor $R_{\mu \nu}$ of the four dimensional spacetime metric
$g_{\mu \nu}$ contracted with the tangent vector of a null geodesic.
The non-negativity of the Ricci tensor is sufficient to establish a
focussing property of
null geodesics which plays an essential role in their proof.

The strategy to be adopted in this paper is to consider not null geodesics
but rather the timelike paths of particles of mass $m$, scalar charge
$gm$ and electric charge ${{s m} \over {\sqrt{4 \pi G}}}$, in a background
metric
$g_{\mu \nu}$, scalar field $\phi$ and vector field $A_{\mu}$.
These timelike
paths may be regarded as the projection of null geodesics moving in
a suitable auxiliary five dimensional metric $g_{AB}$, $A=1,2,3,4,5$
of signature $+++-+$ with Killing field ${\partial} \over {\partial x^5}$.
It then turns out that as long as the Ricci tensor, $^5R_{AB}$ of the
auxiliary five dimensional metric $^5g_{AB}$ is non-negative when
contracted with a null five-vector $V^A$ then the total mass
$M$, scalar charge $\Sigma$ and electric charge $Q$ must satisfy
the \lq \lq anti-gravity bound\rq \rq
$$
M+g \Sigma \ge{ {s |Q|} \over {\sqrt {4 \pi G}}}
\ .\eqno (1.1)
$$
The non-negative Ricci condition, which is the five dimensional null
convergence condition used by Hawking and Ellis [4], will be satisfied
as long as:
$$
g^2 \le 3\ ; \ \ s^2 \ge 2(g^2-1)\ ; \ \ s^2 \le 1+g^2 \ .
\eqno (1.2)
$$
Thus if $0 \le g^2 \le 1$ it suffices that
$$
0 \le s \le \sqrt{1+g^2}\ ,  \eqno(1.3)
$$
if $ 1 \le g^2 \le 3$ it suffices that
$$
\sqrt{2(g^2-1) }\le s \le \sqrt {1+g^2} \eqno(1.4)
$$
while if $g^2 > 3 $ the non-negative Ricci condition will not in general
be satisfied.

The new inequality (1.1) is the same  as that obtained using the
spinorial technique for pure Einstein-Maxwell theory but
differs from it if there are scalar fields.
Recall that
in the absence of scalar  fields (i.e. if $g=0$) and in the absence of
sources
then
the Bogomol'nyi bound [1] is
$$
M \ge { {|Q|} \over {\sqrt {4 \pi G}}}
\eqno (1.5)
$$
and saturation implies that the background is supersymmetric in that it
admits Killing spinors when thought of as a solution of the $N=2$
supergravity theory.

If sources are present then, as pointed out by Sparling and Moreschi [5]
a straightforward modification of the spinorial argument shows that if the
sources have a local energy density $T_{{\overline 0} \, {\overline 0}}$
to charge density $|J _{\overline 0}|$ bounded by
$$
T_{{\overline 0}\, {\overline 0}}  / |J _{\overline 0}| \ge {s \over
 {\sqrt{4 \pi G}}}
\eqno (1.6)
$$
with $0 \le s \le 1$ then
$$
M \ge { {s |Q|} \over {\sqrt {4 \pi G}}} \ .
\eqno (1.7)
$$

If on the other hand the scalar fields are present
but we stick to the case that there are no sources, i.e. if $g \ne 0$
then the spinorial technique gives the bound
$$
M \ge { {|Q|} \over {\sqrt{ 4 \pi  G (1+ g^2)}}} \ .
\eqno (1.8)
$$
Finally if sources are present which satisfy
$$
T_{{\overline 0}\, {\overline 0}}  / |J _{\overline 0}| \ge {se^{-g\kappa\phi}
\over{\sqrt{4 \pi G}}}
\eqno (1.9)
$$
then the spinorial
argument yields (1.7) as long as:
$$
0 \le s \le {1 \over {\sqrt {1+g^2}}} \ .   \eqno(1.10)
$$

It seems clear therefore that since (1.1) and (1.7) do not coincide our new
inequality is giving us some
independent information from that provided by the spinorial method.

\beginsection 2. The Four Dimensional Field Equations

We shall consider theories, possibly with sources, whose field equations
are derivable from an action of the form
$$
\int \sqrt {-g} d ^4 x \Bigl( { R \over {16 \pi G}} - {1 \over 4} \exp (2 g
 \phi {\sqrt{4 \pi G}})
F_{\mu \nu} F ^{\mu \nu} - {1 \over 2} g^{\mu \nu} \partial _{\mu} \phi
 \partial _{\nu}
\phi   \Bigr)
\eqno (2.1)
$$
where $F_{\mu \nu} = \partial _\mu A _ \nu - \partial _\nu A _\mu$ and $g$ is
a dimensionless constant which takes different values for different theories.
For example $g=\surd 3$ for standard Kaluza-Klein theory, $g={1\over\surd3}$
corresponds to dimensionally reduced Einstein-Maxwell theory from five to four
dimensions [1], and $g=1$ corresponds to a truncation of $N=4$ supergravity
theory and is
related to superstring theory. Of course $g=0$ corresponds to Einstein-Maxwell
theory.
Note that we are using signature $+++-$, rational units for the vector field
and we shall
denote $\sqrt{ 4 \pi G}$ by $\kappa$. The latter choice is most convenient for
our present
purposes but it is by no means universal.

An isolated system has mass $M$ defined in the usual way and electric charge
$Q$
 and scalar charge $\Sigma$ defined by
$$
A_0 \sim {Q \over {4 \pi r}}\ , \eqno(2.2)
$$
$$
\phi \sim {G \over \kappa} {\Sigma \over r}  \eqno(2.3)
$$
two such systems with masses $M_1$ and $M_2$, scalar charges $\Sigma _1 $ and
$ \Sigma _2$ and electric charges $Q_1$ and $Q_2$ will experience a net
 attraction of
$$
\Bigl( G M_1 M_2 + G \Sigma _1 \Sigma _2 - {{Q _1 Q _2} \over {4 \pi }}
\Bigr )
{ 1 \over r^2} \ . \eqno(2.4)
$$
The field equations obtained from varying the action are (with appropriate
 additional sources $T_{\mu \nu}$, $J$ and $J_\mu$ )
$$
R _{\mu \nu}= 2 \kappa ^2 \exp (2 g \kappa \phi)
\left( F_{\mu \beta} F_ \nu ^{\ \beta} - {1 \over 4} g_{\mu \nu}
F_{\alpha \beta} F^{\alpha \beta} \right)
+ 2 \kappa^2 \partial _\mu \phi \partial _\nu \phi
$$
$$
+2 \kappa ^2 \left( T _{\mu \nu} -{1\over 2} g_{\mu \nu } T_\alpha^{\ \alpha}
 \right),  \eqno(2.5)
$$
$$
\nabla _\mu ( \exp( 2 g \kappa \phi) F_\alpha  ^{\  \mu} ) =
 J _ \alpha \eqno(2.6)
$$
and
$$
\nabla_\mu(\nabla^\mu\phi - g\kappa e^{2g\kappa\phi}F^{\mu\nu}A_\nu)=
\nabla ^2 \phi - {{ g \kappa } \over 2 }  \exp (2g \kappa \phi)
F _{\mu \nu} F ^ {\mu \nu} =-\kappa J\ .  \eqno(2.7)
$$

The quantities $e^{2g\kappa\phi}F^{\mu\nu}$ and
$(\nabla^\mu\phi - g\kappa e^{2g\kappa\phi}F^{\mu\nu}A_\nu)$ are, in the
absence of the additional sources, the conserved N\"other currents associated
with the transformations
$$A_\mu\mapsto A_\mu+\delta A_\mu\eqno(2.8)$$
and
$$\eqalign{\phi&\mapsto \phi+\delta\phi \cr
A_\mu&\mapsto A_\mu e^{-g\kappa\delta\phi}}\eqno(2.9)
$$
respectively.

{}From the point of view of the Maxwell  field the dilaton field behaves
as a dielectric constant
$$
\epsilon = \exp (2 g \kappa \phi) \eqno(2.10)
$$
and magnetic permeability
$$
\mu = \exp (-2 g \kappa \phi). \eqno(2.11)
$$
Note that the product
$\epsilon \mu$ is unity so the local speed of light is still unity, which
is consistent with local Lorentz invariance.
However now  because empty space behaves in the presence of a dilaton field
a
little like a material medium
one must distinguish the electric field strength $E_i=F_{4i}$
from the divergence-free electric displacement
$D_i = \epsilon  E_i = \exp(2 g \kappa \phi) E_i$ and
divergence-free magnetic induction $B_i= {1 \over 2}
\epsilon_i ^{\ jk} F_{jk}$
from the magnetic field strength
$H_i=\mu ^{-1} B_i = \exp(2 g \kappa \phi) {1 \over 2}
\epsilon_i ^{\ jk} F_{jk}$.
The contribution of the Maxwell field to the local energy
density is thus given by:
$$
{1 \over 2} \bigl( {\bf E}. {\bf D} + {\bf B}.{\bf H}\bigr)\ . \eqno(2.12)
$$

\beginsection {3. The Five Dimensional Metric}

Let us now consider the metric defined on $M_5 = R \times M_4$, where $M_4$ is
the physical spacetime manifold.
$$
ds^2 _5 = e ^{2 \nu} ( dx^5 + s \kappa A _ \mu d x ^ \mu ) ^2 + e ^ {2 \chi} g
_
 {\alpha \beta} dx ^ \alpha dx^ \beta  \eqno(3.1)
$$
where $\nu$ and $\chi$ are scalar fields to be specified later.

Thus the five dimensional metric is a twisted warped product.
Note that in contrast to the standard Kaulza-Klein approach we are
{\sl not} identifying the fifth coordinate $x^5$. Moreover, we will insist that
$e^{2\nu}$ and $e^{2\chi}$ never vanish so that both the five dimensional
metric $g_{AB}$ and the four dimensional metric $g_{\mu\nu}$ are regular
outside any event horizons. In this way we exclude possible counter-examples
involving metrics which while regular in five dimensions are not regular as
four dimensional metrics [6],[7].

Null geodesics of the metric satisfy
$$
{{dx^A} \over {d \lambda}}   = g^{AB}
{ {\partial {\cal S}} \over {\partial x^B}} \eqno (3.2)
$$
where ${\cal S}$ is a solution of the five dimensional
Hamilton-Jacobi equation
$$
g^{AB}
{ {\partial {\cal S}} \over {\partial x^A}   }
{ {\partial {\cal S}} \over {\partial x^B}   } =0\ . \eqno(3.3)
$$
If we set
$$
{\cal S} = {e \over \kappa} x^5 + S \eqno (3.4)
$$
we find that $S$ satisfies the four dimensional Hamilton-Jacobi equation
$$
{g^{\alpha \beta}
\left({{\partial { S}} \over {\partial x^ \alpha}}-seA_\alpha\right)
\left({{\partial { S}} \over {\partial x^ \beta}}-seA_ \beta\right) }
= - {{e^2 } \over {\kappa ^2}} \exp2(\chi -\nu)  \eqno (3.5)
$$
while (3.2) becomes
$$
{{d x^5} \over {d \lambda}} = {e \over \kappa} e ^{-2 \nu}
-s \kappa  e^ {-2 \chi}  g^{\alpha \beta} \left( {{\partial S} \over { \partial
 x ^ \alpha }} - s e A_\alpha \right) A _ \beta  \eqno(3.6)
$$
and
$$
{{d x ^ {\alpha}} \over {d \lambda }} = g ^{ \alpha \beta} e^{-2 \chi}
\left ( { {\partial { S}} \over {\partial x^ \beta} }
 -seA_ \beta \right). \eqno(3.7)
$$

The interpretation of these equations is that the projection of null
geodesics into four dimensions gives the world line of a particle of mass
$$
m = {e \over \kappa}  \eqno(3.8)
$$
and charge
$$
q=se   \eqno(3.9)
$$
with a coupling to the scalar field $(\chi - \nu)$.

If one now considers a four-metric $g_{\mu \nu}$ which, in a quasi-Cartesian
coordinate system, to order $ 1 \over r$ is given by
$$
ds^2 \sim -\left(1-{{2GM} \over r}\right) dt^2 + \left(1+{{2GM} \over r}\right)
\left( dx^2 + dy^2 + dz^2 \right),  \eqno(3.10)
$$
$$
r^2=x^2+y^2+z^2. \eqno(3.11)
$$
with (again to order $1 \over r$)
$$
A_\mu dx ^ \mu \sim {Q \over {4 \pi r}} dt \eqno (3.12)
$$
and
$$ \chi \sim {{\chi _ 0} \over r} \ \ ,\quad \nu \sim {{\nu _0} \over r}
\ .\eqno(3.13)
$$

We wish to solve the Hamilton-Jacobi equation (3.3) at large impact parameter
for a light ray
in the five dimensional metric, one readily finds that the solution (upto
an overall scale is given by)
$$
S=-t+x^5\cos\gamma+z\sin\gamma + 2G\hat M{\rm cosec\,}\gamma\log\left(r+z\over
b
\right)+{\cal E} \eqno(3.14)
$$
where the coordinates have been chosen so that the tangent vector to the null
geodesic is asymptotic to $\bf K$,
$$
{\bf K}=\partial_t+\sin\gamma\partial_z+\cos\gamma\partial_5 \eqno(3.15)
$$
with $\sin\gamma\neq 0$. The impact parameter, $b$, is given by
$$
b^2=x^2+y^2\eqno(3.16)
$$
and ${\cal E}$ is small in the sense that $\partial_b{\cal E},\partial_z{\cal
E}
,\partial_t{\cal E},\partial_5{\cal E} ={\cal O}({1\over b}) $, ${\bf K\cal
E}={
\cal O}({1\over
b^2})$. The value of $\hat M$ is given
by
$$
2G\hat M=2GM-\cos^2\gamma (GM+\chi_0-\nu_0)+{s\kappa Q\over 4\pi}\cos\gamma
\ .\eqno(3.17)
$$
We are considering the case $\sin\gamma\neq 0$ throughout, as for any non-zero
value of $\sin\gamma$  one expects that for large enough impact parameter the
timelike geodesic in four dimensions will be approximated to zeroth order by
the
corresponding timelike geodesic in Minkowski space. In this prescription it is
only the
Schapiro time delay/advance that we wish to take account of and not the bending
of the geodesic in the spacetime, consequently the geodesic is described by a
straight line  in the quasi-Cartesian coordinate system.

Let us calculate $\tau$ the `time of flight' as defined by Penrose, Sorkin
and
Woolgar. We are interested in the Schapiro time delay/advance at large impact
parameter. One evaluates the change in $S$ along a finite section of the
geodesic from  $z_0$ to $z_1$ say, at impact parameter $b$. We are only
interested in the dependence on $b$ and so one subtracts off a similar
contribution at a fixed (large) value of the impact parameter, which we will
call $b_0$. Letting $z_0$ and $z_1$ tend to the initial and final endpoints of
the geodesic (so $z_0\rightarrow -\infty$ and $z_1\rightarrow\infty$ if
$\sin\gamma>0$ whilst if $\sin\gamma<0$, $z_0\rightarrow \infty$ and
$z_1\rightarrow-\infty$) one finds
$$
\tau=-4G\hat M|{\rm cosec\,}\gamma|\log\left(b\over b_0\right)+{\cal O}(1)\ ,
\eqno(3.18)
$$
where ${\cal O}(1)$ signifies bounded as $b\rightarrow\infty$. The conclusion
is that for $\hat M<0$ the time of flight can be made arbitrarily large by
venturing to large enough impact parameter. In other words, there exist a
fastest null curve $\zeta$, in the sense that it minimizes $\tau$.

On the other hand the theorem
proved in [3] shows,
provided that any singularities are inside event horizons,
that $\zeta$ may be taken to be geodesic, and more
especially it lies on the boundary of the causal future of some point on the
appropriate generator of past null infinity, ${\cal I}^-$.
This theorem implies that it cannot have any
conjugate points as no geodesic with conjugate points can stay on the boundary
of the causal future of some point for more than a finite affine length. We
conclude that $\hat M\ge 0$, for all $\sin\gamma \neq 0$,
provided that every null geodesic develops conjugate points (and hence gives a
contradiction to the construction of such a fastest geodesic). The property
that every null geodesic develops conjugate points
is implied by the null convergence condition in the following sections,
together
with an appeal to the generic condition,
$V_{[A} \,^5 R_{B]CD[E}V_{F]}V^C V^D \neq 0$  somewhere on
every null geodesic with tangent vector $V^A$ [4].
The purpose of this condition is to force a congruence of null rays to
start to converge, i.e. the expansion can be made negative and hence by
Raychaudhuri's
equation for null geodesics we can conclude the existence of conjugate points
if we
assume  the non-negativity of the Ricci tensor. The generic condition is not
satisfied by
all spacetimes but any spacetime is arbitrarily close in a certain sense to
one that does.
Our inequalities will therefore hold for spacetimes whether or not they
actually obey
this condition. The null convergence condition guarantees the existence of
conjugate
points and has been employed particularly in relationship to the singularity
theorems of
Penrose, Hawking and others. This requirement seems to require us to relate the
quantities $\chi$ and $\nu$ as follows: $$ \nu= - 2 \chi\ , \eqno(3.19) $$
$$ 3 \nu = 2 g
\kappa \phi \ . \eqno(3.20) $$
By looking at the inequality $\hat M\ge0$ when we take the limits
$\gamma\rightarrow
0,\pi$ one establishes $$ M + g \Sigma \ge {{ |sQ|} \over {\sqrt{4 \pi G}}}\ .
\eqno(3.21) $$

\beginsection {4. The Five Dimensional Ricci Tensor}

In what follows we shall use an orthonormal frame
$$
{\bf E}^A = \left( {\bf E} ^5, e^\chi {\bf e}^\alpha \right)  \eqno(4.1)
$$
where ${\bf e}^\alpha $ is an orthonormal frame with respect to the metric
$g_{\alpha \beta}$. We shall denote by $5$ the ${\bf  E}^5$ component and by
$\alpha$ the component with respect to ${\bf e}^ \alpha$. In an orthonormal
frame and with such an understanding the components of the five dimensional
Ricci tensor $^5R_{AB}$ of the metric $g_{AB}$ are

$$
^5 R_{\alpha \beta} = ^4 R _{\alpha \beta}
-2\nabla _\alpha \nabla _ \beta \chi
- g _ {\alpha \beta} \nabla ^2 \chi
+ 2 \nabla _\alpha  \chi \nabla _\beta  \chi
$$
$$
-2 g_{\alpha \beta} \nabla _ \mu \chi \nabla ^ \mu \chi
- \nabla _\alpha \nu \nabla _\beta \nu
- \nabla _ \alpha \nabla _ \beta \nu
+ \nabla _ \alpha \nu \nabla _\beta \chi
+ \nabla _ \beta \nu \nabla _\alpha \chi
- g _ {\alpha \beta} \nabla _\mu \chi \nabla ^\mu \nu
$$
$$
- { {s^2 \kappa ^2} \over 2 }
e^{2 \nu} e^{-2 \chi} F_{\alpha \sigma} F_\beta ^{\ \sigma}  \ ,\eqno(4.2)
$$

$$
^5 R _{5 \alpha} = {1 \over 2} s \kappa e ^{-2 \nu} e ^ {-2 \chi}
 \nabla _\lambda ( e ^{3 \nu} F_ \alpha ^{\ \lambda} ) \ ,\eqno(4.3)
$$
$$
^5 R_{55} = {{s^2 \kappa^2} \over 4}   e^{2 \nu}    e^{-4 \chi}
F_{\mu \nu} F ^{\mu \nu}
- e^{-2 \chi } ( \nabla ^2 \nu + \nabla _\mu \nu \nabla ^ \mu \nu
+ 2 \nabla _\mu \chi
\nabla ^\mu \nu)\  ,\eqno(4.4)
$$

If one compares (4.2) - (4.4) with the field equations (2.5) - (2.7)
it becomes clear that a good choice
(and possibly the only choice, if one is to
 eliminate the second derivatives over whose sign one would otherwise have no
control) is $$ 3 \nu = 2 g \kappa \phi\ ,\eqno(4.5)
$$
$$
\nu + 2 \chi =0\ . \eqno(4.6)
$$

This gives
$$
M+g \Sigma \ge{ {s |Q|} \over {\sqrt {4 \pi G}}}\ . \eqno(4.7)
$$

Given these choices for $\nu$ and $\chi$ the components of the
five dimensional Ricci tensor are given by
$$
^5R_{\alpha 5} ={1\over2}s\kappa e^{-\nu}J_{\alpha} \  ,\eqno(4.8)
$$
$$
^5R_{55} = \kappa^2 e^{4 \nu} F_{\mu \nu} F^{\mu \nu} \left({{s^2} \over 4} -
{{
g^2}
 \over 3}\right)+{2g\kappa^2\over3}e^\nu J ,\eqno (4.9)
$$
and
$$
^5 R_{\mu \nu}= 2 \kappa^2 \left(1-{{g^2} \over 3}\right) \partial_ \mu  \phi
\partial _
\nu \phi
+ 2 \kappa ^2 e^{3\nu}\left( F_{\mu \beta} F_\nu ^{\ \beta} - {1 \over 4}
 g_{\mu \nu}
F_{\alpha \beta} F^{\alpha \beta} \right)
$$
$$
- {{s^2 \kappa^2} \over 2} e^{3 \nu} F_{\mu \beta} F_ \nu ^{\  \beta}
+ {{g^2 \kappa ^2} \over 6}e^{3\nu} g_{\mu \nu} F_{\alpha \beta} F^{\alpha
\beta}+2\kappa^2\left(T_{\mu\nu}-{1\over2}g_{\mu\nu}T_\alpha^{\ \alpha}\right)
\ .\eqno(4.10)
$$
As a check we note that the standard five dimensional
theory without additional sources corresponds to the values
$$
g^2=3  \eqno(4.11)
$$
and
$$
s=2\ .  \eqno (4.12)
$$
In this case we have
$$^5 R _{AB} =0\ . \eqno(4.13)
$$

\beginsection {5. The Null Convergence Condition in Five Dimensions}

We must now calculate
$$
^5 R _{AB} { {dx^A} \over {d \lambda}} { {dx^B} \over {d \lambda}} \eqno(5.1)
$$
and find under what conditions it is non-negative.
The lightlike vector $V^A = {{dx^A} \over {d \lambda}}$ has components
$$
{\bf V} = V_5 {\bf E}^5 + V_\alpha {\bf e}^\alpha  ,\eqno(5.2)
$$
so that
$$
\left( V_5 \right)^2 = - e^{-2 \chi} g^{\alpha \beta} V_\alpha V_ \beta\ .
\eqno
(5.3)
$$
One therefore has
$$
^5R _{AB} V^A V^B = 2\kappa^2
e^{2\nu}\left(1-{g^2\over3}\right)\left(V^\mu\partial_\mu\phi\right)^2
$$
$$ + \kappa^2e^\nu(V_5)^2\left[\left(1-g^2+{s^2\over2}\right) {\bf B.H} +
\left(1+g^2-s^2\right){\bf E.D}\right]
$$
$$
+2\kappa^2e^{2\nu}\left(T_{\mu\nu}-{1\over2}g_{\mu\nu}T_\alpha^{\
\alpha}\right)V^\mu V^\nu
+{2\kappa^2\over3}(V_5)^2e^\nu gJ
+s\kappa V_5V^\mu J_\mu \eqno(5.4)
$$
where we define
$$
\eqalign{E_\nu&={e^{-\chi}V^\mu\over|V_5|} F_{\mu\nu}\, ,\cr
         B_\mu&={1\over2}\epsilon_{\mu\nu\rho\sigma}{e^{-\chi}V^\nu\over|V_5|}
F^{\rho\sigma}\ .}\eqno(5.5)
$$
and
$$
\eqalign{{\bf D}&=\epsilon{\bf E}=e^{2g\kappa\phi}{\bf E}\ ,\cr
         {\bf H}&=\mu^{-1}{\bf B}=e^{2g\kappa\phi}{\bf B}\ .}\eqno(5.6)
$$
Hence, (5.1) will be non-negative provided we impose the conditions,
$$
s^2\le 1+g^2 \hbox{\qquad and \qquad} s^2\ge 2(g^2-1)\ ,\eqno(5.7)
$$
and the sources satisfy
$$
J\ge0\, ,\eqno(5.8)
$$
$$
\left(T_{\mu\nu}-{1\over2}g_{\mu\nu}T_\alpha^{\ \alpha}\right)\hat V^\mu
\hat V^\nu\ge
e^{-g\kappa\phi}\left|{s \hat V^\mu J_\mu\over2\kappa}\right|\ .\eqno(5.9)
$$
Here $\hat V$ is a unit timelike vector in four dimensions. These conditions
are sufficient for the theorem to hold, though all we require is the
non-negativity of the right hand side of equation (5.4). Equation (5.9) relates
the gravitating energy (in the sense that it is the source for the Ricci tensor
and hence it appears as the appropriate density
in the Poisson equation in the Newtonian limit)
to the local energy density of the vector field, and
should be compared to (1.9) which is the appropriate condition for the
spinorial technique to apply. Indeed one notices that
for a pressure-free fluid the two equations (1.9) and (5.9) coincide.

\beginsection {6. Comparison with Black Hole Solutions}

It is interesting to compare our inequalities with the explicit
spherically symmetric solutions of the field equations.

The four-metric is
$$
ds^2=-\left(1 -{{r_+} \over r}\right)\left(1 - {{r_-} \over r}\right)^{{1-g^2}
 \over {1+g^2}} dt^2
+\left(1 -{{r_+} \over r}\right)^{-1}\left(1 - {{r_-} \over r}\right)^{{g^2-1}
 \over {1+g^2}} dr^2$$
$$
+r^2 \left(1-{{r_-} \over r}\right)^{{2g^2} \over {1+g^2}} \left(d \theta ^2
+ \sin ^2 \theta d \phi ^2\right) \eqno(6.1)
$$
with scalar field
$$
e^{\kappa \phi} = \left(1- {{r_-} \over r}\right)^ {{-g} \over {1+g^2}}
\eqno(6.
2)
$$
and Maxwell field
$$
F= e^ {-2g \kappa \phi}{{Q dt \wedge dr} \over {4 \pi r^2 \left(1-{{r_-}
\over r}\right)^{{2g^2} \over {1+g^2}}}}\  .\eqno(6.3)
$$
One has
$$
GM= {1 \over 2}\left(r_+ + r_- {{1-g^2} \over {1+g^2} } \right), \eqno(6.4)
$$
$$
|Q| \sqrt{ { {G} \over {4 \pi} }} = { \sqrt{r_+ r_-} \over {\sqrt{1+g^2}}}
\eqno
(6.5)
$$
and
$$
G \Sigma = {{gr_ -} \over {1+g^2}}\ . \eqno(6.6)
$$
Whence
$$
G(M + g \Sigma) ={1 \over 2} (r_+ + r_-) \ge \sqrt {r_+ r_-}= G
\sqrt{1+g^2} {|Q| \over {\sqrt{4 \pi G}}} \eqno(6.7)
$$
which is consistent with our general result. Note however that in obtaining
(6.7) for this example we have not needed to restrict the coupling constant $g$
to be less than
$\surd 3$. Thus it seems that while $g^2 \le 3$
is a sufficient condition for the validity of our inequality (1.1)
it may not be
a necessary condition. This opens up the possibility
that one might
be able to extend our proof beyond the the case $g^2 \le 3$ by using
some other judiciously chosen metric.

\beginsection 7. The Extreme Case

Let us consider the analogues of Papapetrou-Majumdar solution in the theory we
have been considering. Gibbons [8] has found the appropriate form of the
metric:
$$
ds^2=-H^{-2\over1+g^2}dt^2+H^{2\over1+g^2}(dx^2+dy^2+dz^2)\, ,\eqno(7.1)
$$
with $H$ a harmonic function in Euclidean space with Cartesian coordinates
$x,y$ and $z$. The dilaton and vector potential are given by
$$
e^{\kappa\phi}=H^{g\over1+g^2}\eqno(7.2)
$$
and
$$
A={dt\over\kappa H\sqrt{1+g^2}}\eqno(7.3)
$$
We wish to relate this solution to a five dimensional metric:
$$
ds_5^{\ 2}=H^\alpha\left(du+{s dt\over H\sqrt{1+g^2}}\right)^2+
H^\beta\left(-H^{-2\over1+g^2}dt^2+H^{2\over1+g^2}(dx^2+dy^2+dz^2)\right)\ .
\eqno(7.4)
$$
We write $u=x^5-t$ in equation (7.4), as the gauge potential in equation (7.3)
does not conform to our usual gauge choice. We demand
$H({\bf x})\rightarrow 1$ as $|{\bf x}|\rightarrow\infty$ as a condition in
order to have an asymptotically flat solution in standard quasi-Cartesian
coordinates.
The choice of $\alpha,\ \beta$ and $s$ will be made in such a way that the
resulting five dimensional metric possesses a lightlike Killing vector
$\partial\over\partial t$. We are therefore led to the conditions:
$$
\eqalign{&s=\sqrt{1+g^2}\,, \cr
&\alpha-\beta={2g^2\over1+g^2}\,.}\eqno(7.5)
$$
On the other hand, comparing (7.4) with our general ansatz for the metric we
write
$$
H^\alpha=e^{2\nu}=e^{4g\kappa\phi\over3}=
H^{{4\over3}{g^2\over1+g^2}}\ .\eqno(7.6)
$$
Hence from (7.5) we deduce that
$$
\alpha+2\beta=0\,,\eqno(7.7)
$$
which is a necessary relationship if we are to eliminate second order
derivatives of the dilaton from the expansion of the Ricci tensor in section 4,
and hence to impose the null convergence condition on the Ricci tensor.
The lightlike Killing vector $K$ has components
$$
K_\mu=H^{\alpha-1}\nabla_\mu u\, ,\eqno(7.8)
$$
and hence is hypersurface orthogonal. The relevant hypersurfaces being those of
constant $u$.

We now wish to show that (7.1) saturates the bound, in the sense that $\hat
M\rightarrow0$ as $\gamma\rightarrow0$. Consider the case $\gamma=0$ exactly.
The corresponding null geodesics are null generators of the null surface
$u={\rm constant}$. The associated
exact solution of the Hamilton-Jacobi equation is given
(upto scalar multiplication and the addition of a constant) by
$$
S=u=-t+x^5,\eqno(7.9)
$$
independent of the value of the impact parameter. If this is compared with
the $\gamma\rightarrow 0$ limit of (3.14),
one concludes that (7.1) is a solution that saturates the bound, i.e.
$\hat M=0$ with $\gamma=0$ so that
$$
M+g\Sigma={-sQ\over\kappa}\,.\eqno(7.10)
$$
Equation (7.10) may also be shown by direct computation. Set
$$
H=1+\sum_{i=1}^n {G\mu_i\over|{\bf x}-{\bf x_i}|}\ ,
$$
representing $n$ isolated black holes of ``strengths'' $(\mu_i)_{i=1}^n$
at locations $({\bf x_i})_{i=1}^n$ in equilibrium.
One calculates the mass, scalar and electric charges:
$$
\eqalignno{&M={1\over1+g^2}\sum_{i=1}^n \mu_i\ ,&(7.11)\cr
         &Q={-\kappa\over\sqrt{1+g^2}}\sum_{i=1}^n \mu_i &(7.12)}
$$
and
$$
         \Sigma={g\over1+g^2}\sum_{i=1}^n \mu_i\ .\eqno(7.13)
$$
Together with $s=\sqrt{1+g^2}$, we verify that (7.10) is indeed satisfied.

\beginsection 8. Summary and Conclusion

In this paper we have considered four dimensional spacetimes, Maxwell field
$A_\mu$ and scalar field $\phi$ with associated five dimensional metric:
$$
ds^2=e^{4g\kappa\phi/3}\left(dx^5+s\kappa A_\mu
dx^\mu\right)^2+e^{-2g\kappa\phi/3}g_{\alpha\beta}dx^\alpha dx^\beta\eqno(8.1)
$$
where $g$ and $s$ are dimensionless constants. Assuming that any
singularities are contained within an event horizon and that the five
dimensional Ricci tensor $^5R_{AB}$ obeys the non-negative Ricci condition:
$$
^5R_{AB}V^A V^B\ge0\eqno(8.2)
$$
for all null five-vectors $V^A$, we have shown that the mass, electric and
scalar charges satisfy the relationship:
$$
M+g\Sigma\ge{s|Q|\over\sqrt{4\pi G}}\ .\eqno(8.3)
$$
If we impose the Einstein equations and the equations of motion for the dilaton
and Maxwell fields arising form the Lagrangian (2.1), we find that (8.2) may be
satisfied provided the additional sources $J$, $J_\mu$ and $T_{\mu\nu}$ obey
$$
J\ge0\, ,\eqno(8.4)
$$
$$
\left(T_{\mu\nu}-{1\over2}g_{\mu\nu}T_\alpha^{\ \alpha}\right)\hat V^\mu
\hat V^\nu\ge
e^{-g\kappa\phi}\left|{s \hat V^\mu J_\mu\over2\kappa}\right|\eqno(8.5)
$$
for all timelike unit four-vectors $\hat V^\mu$
and the quantities $s$ and $g$ satisfy the following:
$$
2(g^2-1)\le s^2\le 1+g^2,\eqno(8.5)
$$
for which we require $g\le\surd3$.

Our proof is valid in the full non-linear theory but it is illuminating to see
how it follows almost trivially in the linearized theory. In De Donder gauge
linear theory gives
$$
-{1\over2}\nabla_C\nabla^C h_{AB}=R_{AB}\eqno(8.6)
$$
where $g_{AB}=\eta_{AB}+h_{AB}$. If the metric is independent of $t$ and $x^5$
we have from (8.6) near infinity
$$
h_{AB}={h_{AB}^0\over r}+{\cal O}\left({1\over r^2}\right)\eqno(8.7)
$$
where
$$
h_{AB}^0={1\over2\pi}\int_{{\bf R}^3} dx^3\,R_{AB}\ .\eqno(8.8)
$$
At the linear level the null convergence condition on $^5R_{AB}$ implies that
$h_{AB}^0$ is non-negative when contracted with any constant lightlike vector.
Because
$$
\left(h_{AB}^0\right)=G\pmatrix{\ddots&&&\vdots&\vdots\cr
      &2\left(M-{g\Sigma\over3}\right)&&0&0\cr
  &&\ddots&\vdots&\vdots\cr
 \dots&0&\dots&2M+{2g\Sigma\over3}&{sQ\over\sqrt{4\pi G}}\cr
\dots&0&\dots&{sQ\over\sqrt{4\pi G}}&{4g\Sigma\over3}\cr}
\eqno(8.9)
$$
where the order of rows and columns in (8.9) is $(i,4,5)$, inequality (8.3)
follows by considering the five-vector $(V^A)=({\bf0},1,\pm1)$. If the metric
is independent of $x^5$ but depends on time $t$ we expect the same conclusion
to follow from linear theory. We have not checked this point explicitly since
our stronger non-linear argument does not require the metric to be
time-independent.

{\parindent=0 pt
\beginsection{References}

[1] G.W. Gibbons, D.A. Kastor, L.A.J. London, P.K. Townsend and J.H. Traschen,

{(\it to be published)}

[2] E. Witten, Comm. Math. Phys. {\bf80} 381-402 (1981)

[3] R. Penrose, R.D. Sorkin and E. Woolgar,
{\it (Submitted to Comm. Math. Phys.)}

--- gr-qc 9301015

[4] S.W. Hawking and G.F.R. Ellis, `The Large Scale Structure of
Space-time'~(CUP)~(1973)

[5] O.M. Moreschi and G.A.J. Sparling, Comm. Math. Phys. {\bf95} 113-120 (1984)

[6] E. Witten, Nucl. Phys. {\bf B195} 481-492 (1982)

[7] D. Brill and H. Pfister, Phys. Lett. {\bf B228} 359-362 (1989)

[8] G.W. Gibbons, Nucl. Phys. {\bf B207} 337-349 (1982)

}

\bye